\documentclass[10pt,final,conference]{IEEEtran}%
\usepackage{amsmath}
\usepackage{amsfonts}
\usepackage{amssymb}
\usepackage{graphicx}%
\newtheorem{theorem}{Theorem}

\newcommand{\commentout}[1]{}

\begin{document}

\title{Regret and Jeffreys Integrals in Exp. Families}
\author{Peter Gr\"{u}nwald and Peter Harremo\"{e}s\linebreak\\{\small Centrum Wiskunde \& Informatica\linebreak}\\{\small Amsterdam, The Netherlands\linebreak}\\{\small Emails: Peter.Grunwald@cwi.nl and Peter.Harremoes@cwi.nl}}
\maketitle

\section{Preliminaries}

Let $\{P_{\beta}\mid\beta\in\Gamma^{\text{can}}\}$ be a 1-dimensional
exponential family given in a canonical parameterization,
\begin{equation}
\frac{dP_{\beta}}{dQ}=\frac{1}{Z(\beta)}e^{\beta x},\label{eq:expdef}%
\end{equation}
where $Z$ is the partition function $Z(\beta)=\int\exp(\beta x)~dQx$, and
$\Gamma^{\text{can}}:=\{\beta\mid Z(\beta)<\infty\}$ is the \emph{canonical
parameter space}. We let $\beta_{\sup}=\sup\{\beta\mid\beta\in\Gamma
^{\text{can}}\}$, and $\beta_{\inf}$ likewise. 

The elements of the exponential family are also parametrized by their mean
value $\mu$. We write $\mu_{\beta}$ for the mean value corresponding to the
canonical parameter $\beta$ and $\beta_{\mu}$ for the canonical parameter
corresponding to the mean value $\mu.$ For any $x$ the maximum likelihood
distribution is $P_{\beta_{x}}.$ The \emph{Shtarkov integral} $S$ is defined
as%
\begin{equation}
S=\int\frac{1}{Z(\beta_{x})}e^{\beta_{x}x}~dQx.\label{eq:shtarkovintegral}%
\end{equation}

The variance function $V$ is the function that maps $\mu\in M$ into the
variance of $P^{\mu}.$ The Fisher information of an exponential family in its
canonical parametrization is $I_{\beta}=V\left(  \mu_{\beta}\right)  $ and the
Fisher information of the exponential family in its mean value parametrization
is $I^{\mu}=\left(  V\left(  \mu\right)  \right)  ^{-1}.$ The \emph{Jeffreys
integral} $J$ is defined as%
\begin{equation}
J=\int_{\Gamma^{\text{can}}}I_{\beta}^{1/2}~d\beta=\int_{M}\left(  I^{\mu
}\right)  ^{1/2}~d\mu.~\label{eq:jeffreysintegral}%
\end{equation}
More on Fisher information can be found in \cite{Grunwald2007}.

As first established by \cite{Rissanen1996}, if the parameter space is
restricted to a compact subset of the interior of the parameter space with
non-empty interior (called an \emph{ineccsi\/} set in \cite{Grunwald2007}),
then the minimax regret is finite and equal to the logarithm of the Shtarkov
integral, which in turn is equal to
\begin{equation}
\frac{1}{2}\log\frac{n}{2\pi}+\log J+o(1).\label{eq:asymp}%
\end{equation}
It thus becomes quite relevant to investigate whether the same thing still
holds if the parameter spaces are \emph{not\/} restricted to an ineccsi set.
Whether or not this is so is discussed at length and posed as an open problem
in \cite[Chapter 11, Section 11.1]{Grunwald2007}. 

\section{Results}

\begin{theorem}
\label{thm:start} For a 1-dimensional left-truncated exponential family, the
following statements are all equivalent:

\begin{enumerate}
\item The Shtarkov integral is finite.

\item The minimax individual-sequence regret is finite.

\item The minimax expected redundancy is finite.

\item The exponential family has a dominating distribution $Q_{\text{dom}}$ in
terms of information divergence, i.e. $\sup_{\beta\in\Gamma^{\text{can}}%
}D(P_{\beta}\Vert Q_{\text{dom}})<\infty.$

\item There is distribution $P_{\beta}$ with $\beta\in\Gamma^{\text{can}} $
that dominates the exponential family in terms of information divergence.

\item The information channel $\beta\rightarrow P_{\beta}$ has finite capacity.

\item There exists $\beta_{0}\in\Gamma^{\text{can}}$ such that
\[
\lim_{\beta\uparrow\beta_{\sup}}D(\beta_{0}\Vert\beta)<\infty
\ \ \text{\textrm{or}\emph{\/}}\lim_{\beta\uparrow\beta_{\sup}}D(\beta
\Vert\beta_{0})<\infty.
\]

\end{enumerate}
\end{theorem}

\ \noindent Most of the equivalences between (1)--(6) are quite
straightforward. The surprising part is the fact that statements (1)--(6) are
also equivalent to $(7)$. 

\begin{theorem}
\label{thm:infjef} Let $(\Gamma^{\text{can}}_{0},Q)$ represent a
left-truncated exponential family. If the Shtarkov integral is infinite, then
the Jeffreys integral is infinite.
\end{theorem}

The converse does not hold in general. 

\begin{theorem}
\label{prop:betacasejeffinite} Let $(\Gamma_{0}^{\text{can}},Q)$ represent a
left-truncated exponential family such that $\beta_{\sup}=0$ and $Q$ admits a
density $q$ either with respect to Lebesgue measure or counting measure. If
$q(x)=O(1/x^{1+\alpha})$ for some $\alpha>0$, then the Jeffreys integral
$\int_{\beta}^{0}I(\gamma)^{1/2}~d\gamma$ is finite.
\end{theorem}

In most cases finite Shtarkov implies finite Jeffreys.

\begin{theorem}
Let $Q$ be a measure on the real line with support $I$. Assume that $\mu
_{\inf}$ is the left end point of $I$. If $Q$ has density $f\left(  x\right)
=\left(  x-\mu_{\inf}\right)  ^{\gamma-1}g\left(  x\right)  $ in an interval
just to the right of $a$ where $g$ is an analytic function and $g\left(
\mu_{\inf}\right)  >0$ then the left end of the interval $I$ gives a finite
contribution to Jeffrey's integral if and only if $Q$ has a point mass in $a$.
\end{theorem}

\label{ExpCauchy}If $Y$ is a Cauchy distributed random variable then
$X=\exp\left(  Y\right)  $ has density
\[
\frac{1}{\pi}\frac{1}{x\left(  1+\log^{2}\left(  x\right)  \right)  }.
\]
A probability measure $Q$ is defined as a $1/2$ and $1/2$ mixture of a point
mass in 0 and an exponentiated Cauchy distribution. The exponential family
based on $Q$ has redundancy upper bounded by $1$ bit but the Jeffreys integral
is infinite.

For exponential families in more dimensions the analysis becomes more more
involved and one may even have exponential families with finite redundancy and
infinite regret.

\bibliographystyle{ieeetr}
\bibliography{database}

\end{document}